\documentstyle[12pt,dinA4]{article}

\title{Dilepton Production \\
       in Nucleon-Nucleon Interactions
\footnote{Supported by BMFT and GSI Darmstadt}}
\author{M. Sch\"afer\thanks{Part of dissertation of M. Sch\"afer},
        H. C. D\"onges, A. Engel\\
        and U. Mosel  
\\Institut f\"ur Theoretische Physik, Universit\"at Giessen\\
D--35392 Giessen, Germany}

\begin{document}
\pagestyle{empty}

\titlepage

\maketitle

\begin{abstract}
Starting from a realistic one--boson--exchange--model fitted to
the amplitudes of elastic nucleon--nucleon scattering and the process
$NN\rightarrow N\Delta$ we perform a fully relativistic
and gauge invariant calculation for the dilepton production in
nucleon--nucleon collisions, including the important effect of
propagating the $\Delta$--resonance. We compare
the results of our calculations with the latest experimental
data on dilepton production. We also show how to implement
various electromagnetic formfactors for the hadrons in our calculations
without loosing gauge--invariance and discuss their influence on
dilepton spectra.

\end{abstract}

\newpage

\pagestyle{plain}
\section{INTRODUCTION}
Heavy--ion reactions of intermediate energies up to a few
GeV/u offer the unique possibility to produce chunks of nuclear
matter of densities of 2 -- 3 $\rho_0$ \cite{Mosel}. It is
predicted by various models \cite{WeiseWiesbaden} that at these densities
first effects from the restauration of chiral symmetry, one of
the fundamental symmetries of QCD, should appear. In particular,
a lowering of the masses of the vector mesons has been predicted
\cite{Hatsuda} that could possibly be observed in measurements of
dilepton spectra. In ref.\cite{Wolf93} we have investigated the
observable consequences of such changes.

In this paper we study the elementary production process
for dileptons in nucleon--nucleon collisions, since the
quantitative understanding of this process is the natural
prerequisite to an unequivocal determination of the in--medium
effects mentioned above. We also point out that the study of this
process may give fundamental information on the electromagnetic
formfactor of the nucleon in the timelike region around the
vector meson masses that is otherwise not accessible \cite{Mosel}.

For the production
of photons the relevant cross sections were presented
in \cite{Schaefer91}.
Similar calculations were performed for dileptons, but the effect
of resonances was neglected \cite{Haglin89,Schaefer89,Haglin91}.
In ref.\cite{Haglin91} calculations quite similar to our calculations
for the nucleon contribution to the cross section  were
performed but there only back--to--back pairs of the dileptons were
considered, because in this case the phasespace for the nucleons
can be integrated out in an analytic way. The cross section obtained
in this way was
assumed to be also valid for dileptons that are not
back--to--back pairs.

In chapter 2 we will shortly present our general calculational
procedure based on an effective OBE--interaction
where all the parameters are determined by fitting the
elastic nucleon-nucleon scattering data and the NN$\rightarrow $
N$\Delta $ reaction. Using this set of parameters in chapter 3 our
results for dilepton production will be presented.

Since all hadrons have an electromagnetic structure, we have to include
formfactors in our calculation, too.
How it is possible to do this without loosing gauge invariance
is shown in subsection 1 of chapter 3. In the last subsection of chapter 3
we present the results of our
calculations using the parameters found in chapter two. The conclusion
in chapter four will end our paper.

\section{Effective OBE scattering amplitude}
To obtain the desired production cross sections we use
covariant perturbation theory in first order in the
nucleon-nucleon interaction. Particles are simply produced from
the external nucleon lines or, if possible, from
the internal meson lines (Fig. 1). Particle emission between subsequent
NN interaction vertices are omitted in our calculation scheme, but we include
the more important effects of excited $\Delta$-resonances.

To determine all the parameters involved in our calculation we need as
an essential input the nucleon-nucleon interaction. In contrast to some
iterative scheme we directly fit the $T$-matrix by
an effective nucleon-nucleon interaction based on an One-Boson-Exchange Model.
For lab energies up to 400 MeV such fits were performed, using up to 16
mesons \cite{Horowitz85,Horowitz87}.
Here we restrict ourselves to the most essential four mesons for fitting
elastic
nucleon-nucleon scattering data for lab energies in the range from
800 MeV up to 3 GeV. For these four mesons we use the following Hamiltonian:
\begin{eqnarray}
\label{ALLHAMILTONIAN}
H_{int}&=&g_\sigma\bar\psi\psi\sigma+g_\omega\bar\psi\gamma_\mu\psi\omega^\mu
\nonumber\\
 &+&g_\rho\bar\psi\left(\gamma_\mu + \frac{\kappa\partial_\mu}{2m_N}\right)
\vec\tau\psi\vec\rho^\mu\nonumber\\
 &+&ig_\pi\bar\psi\gamma_5\vec\tau\psi\vec\pi
\end{eqnarray}
where $\sigma, \omega, \vec\rho$ and $\vec\pi$ denote the
scalar, vector, isovector-vector and isovector-pseudoscalar
fields , respectively, of the four mesons.

{}From the pion-nucleon interaction the NN potential
\begin{eqnarray}
\label{PIONPOTENTIAL1}
V_\pi(q)&=&-\left(\frac {g_\pi}{2m_N}\right)^2 \vec\tau_1\cdot\vec\tau_2
                                             \nonumber\\
 &\times &\left(\frac{\vec\sigma_1\cdot\vec q\quad\vec\sigma_2\cdot\vec q}
               {m^2_\pi + \vec q^2}\right)
\end{eqnarray}
can be deduced. One feature of this potential is that for $\vec q = 0$,
i.e. for the forward direction in the direct diagrams and the backward
direction in the exchange diagrams,
the scattering amplitude vanishes; this leads to an
unreasonable shape of the elastic scattering cross section. To avoid
this, we first decompose the potential eq.(\ref{PIONPOTENTIAL1})
into the sum of a tensor part and a central part.
After a Fourier-transformation of the central part we get the well-known
Yukawa-interaction
and an additional contact term \cite{WeiseBuch}.
\begin{eqnarray}
\label{PIONRSPACE}
V^C_\pi(r)&=&\frac{m^3_\pi}{12\pi}\left(\frac{g_\pi}{2m_N}\right)^2
\vec\sigma_1\cdot\vec\sigma_2\quad\vec\tau_1\cdot\vec\tau_2\nonumber\\
          &\times&\left(\frac{e^{-m_\pi r}}{m_\pi
r}-\frac{4\pi}{m^3_\pi}\delta(r)
                   \right)
\end{eqnarray}
It is the contact term that causes the unphysical behaviour in the angular
distribution. In nature this term is effectively turned off by the repulsive
hard core of the nucleon--nucleon interaction, but in our scheme
it has to be explicitly subtracted.
This is achieved by adding the following Hamiltonian:
\begin{eqnarray}
\label{CONTACTTERM}
\Delta H_{int}&=&g_a\left(\bar\psi\gamma_5\gamma_\mu\vec\tau\psi\right)
                   \left(\bar\psi\gamma_5\gamma^\mu\vec\tau\psi\right)\\
 g_a&=&-\frac{1}{3}\left(\frac{g_\pi}{2m_N}\right)^2\nonumber
\end{eqnarray}
since
\begin{eqnarray}
\gamma_0\gamma_5\gamma_\mu =\left(-\gamma_5, \vec\sigma\right).
\end{eqnarray}
Eq.(\ref{CONTACTTERM}) describes the interaction of the nucleons via
a very heavy axialvector--isovector meson.
For high bombarding energies
and inelastic nucleon-nucleon processes this interaction generates
additional terms besides the desired $\delta$-force in coordinate
space, but these are of minor importance in comparison to the influence
of the $\delta$-force. Fig. 2 shows clearly the improvement
obtained when using the contact term eq.(\ref{CONTACTTERM}) in
addition to a pion-potential.

To take into account the finite size of the nucleons
we introduce formfactors
\begin{eqnarray}
\label{STRONGFORMFACTOR}
F_i(q^2)=\frac{\Lambda^2_i-m^2_i}{\Lambda^2_i-q^2} ,
\end{eqnarray}
at each strong-interaction vertex where $q$ is the four-momentum and $m$ the
mass of the exchanged meson.

At total we have eight parameters which have to be determined by elastic
nucleon-nucleon scattering data. In an effective interaction scheme
it is not possible to reproduce all the data in the desired
energy range from 800 MeV up to 3 GeV with energy independent
parameters. Therefore we have kept the four cutoffs $\Lambda$ energy
independent and use for each meson a coupling constant which
depends on the total c.m. energy
\begin{eqnarray}
\label{COUPLINGS}
g(\sqrt{s})=g_0 e^{-l\sqrt{s}} \quad .
\end{eqnarray}

With these 12 parameters $(\Lambda, g(0), l)_i$ given in Table 1
we fit the $T$-matrix
to the relevant pp and pn data at three different
laboratory energies 1.73, 2.24 and 3.18 GeV for proton-proton and
proton-neutron scattering. In Fig. 3 we show
the results of our calculations using these parameters
in comparison with the experimental data \cite{Lan}.
Note that the 800 MeV-data (Fig. 3 top) were not included in our
fit procedure. From this comparison we conclude that the elastic
amplitudes are quite well described. We then assume that the
NN--meson vertices determined by the procedure just described are correct
also for half--off--shell processes.

As already mentioned, for particle production in the energy region
we are interested in,
it is necessary to include higher resonances. In this paper we limit
ourselves to the implementation of the $\Delta$.
The Feynman diagrams for the particle production involving the $\Delta$
always look like the process $N + N \rightarrow N + \Delta $ and the
decay of this $\Delta$ into a nucleon and the desired particle (pion or
dilepton
pair).
Due to isospin conservation only isospin-1-mesons can be exchanged;
in our calculation these are the pion and the rho-meson. The corresponding
vertex functions are:
\begin{eqnarray}
\label{NUCLEONDELTAMESON}
-\frac{f_{N\Delta\pi}}{m_\pi}q^\mu\vec T\quad\quad\mbox{for the pion}\\
-i\frac{f_{N\Delta\rho}}{m_\rho} \left(q^\beta\gamma_\beta g_{\mu\alpha}
 - q_\mu\gamma_\alpha\right)\gamma_5\vec T\quad\quad\mbox{for the rho}
\end{eqnarray}
where $q_\mu$ denotes the momentum of the outgoing meson and $\vec T$
the isospin-operator. From the decay
$\Delta\ \rightarrow N + \pi$ we have determined the $N\Delta\pi-$coupling
constant to $f_{N\Delta\pi}=2.13$.

The $\Delta$ propagator \cite{Ber} is :
\begin{equation}
\label{DELTAPROPAGATOR}
G^{\mu\nu}_\Delta (p) =  - \frac{(p_\eta \gamma^\eta + m_\Delta)}{p^2
-m^2_\Delta}
\left(g^{\mu\nu}-\frac{1}{3}\gamma^\mu \gamma^\nu -\frac{2}{3m^2_\Delta}p^\mu
p^\nu
- \frac{1}{3m_\Delta} (\gamma^\mu p^\nu - p^\mu \gamma^\nu )\right)
\end{equation}

Concerning this propagator there is a some confusion in the literature,
ref.\cite{Williams} used $p^2$ in the projector part of the propagator instead
of $m_\Delta^2$.
This, however, leads to an unreasonable behaviour,
since the projector develops a pole for off-shell $\Delta$'s
with $q^2=0$, which shows up as a pole in the differential cross section
for pion production.

The mass of the $\Delta$ in the denominator is modified by an imaginary width
$m_\Delta \rightarrow m_\Delta - {\rm i} \Gamma / 2$ due to the fact
that $\Delta$'s are not stable particles. Nonrelativistic calculations of the
$\Delta$ selfenergy lead to a $k^3$ dependence of
the width \cite{Weise77} where $k$
is the pion momentum in the rest frame of the $\Delta$. We take
\cite{Dmitriev86}
\begin{eqnarray}
\label{DELTAWIDTH}
\Gamma (\mu^2) &=& \Gamma _0 \left[ \frac{k(\mu^2)}{k(m_\Delta^2)} \right]^3
                   \frac{k^2(m_\Delta^2) + \kappa^2}
                        {k^2(\mu^2)      + \kappa^2} \nonumber \\
k(\mu^2) &=& \sqrt{ \frac{ (\mu^2+m_N^2-m_\pi^2)^2}
                         {4\mu^2} - m^2_N }
\end{eqnarray}
where $\mu$ is the mass of an off-shell $\Delta$.
The constant
$\Gamma_0$ is the free width of 0.12 GeV and $\kappa$ is fixed to 0.16 GeV.

As in the case of nucleon-nucleon scattering (\ref{STRONGFORMFACTOR})
we use formfactors for the nucleon--$\Delta$ vertex but now of dipole form
\cite{DeltaForm}
in order to keep the self--energy of the nucleon finite (a monopole
formfactor would not change any of the results reported below if we use
a somewhat smaller cut--off).
\begin{eqnarray}
\label{STRONGDELTAFORMFACTOR}
     F^{*}_i(q^2)=\left( \frac{\Lambda_i^{* 2}-m^2_i}
                                  {\Lambda_i^{* 2} -q^2} \right)^2
                                  \nonumber \\
    i = \pi ,\rho \quad .
\end{eqnarray}
The three new parameters
$f_\rho $, $\Lambda^*_\pi $ and $\Lambda^*_\rho $ are fitted to the existing
data \cite{Dmitriev86} for mass-differential cross sections for
$N + N \rightarrow N + \Delta $ in the energy range 1 - 2.5 GeV.
For the calculation of the $\Delta$-production we take also into account the
decay width of the $\Delta$ by multiplying
\begin{eqnarray}
\label{MASSDISTRIBUTION}
\rho(\mu) = \frac{1}{\pi} \frac{m_\Delta\Gamma(\mu^2)}
            {(\mu^2-m_\Delta^2)^2+m_\Delta^2\Gamma^2(\mu^2)}
\end{eqnarray}
to the cross section obtained from the Feynman diagram.

In a fit to experimental data of $pp \rightarrow n\Delta^{++} $
at 3 energies (970 MeV, 1.48 GeV, 2.02 GeV) we have obtained
\begin{eqnarray}
\label{DELTAPARAMETERS}
  \Lambda^*_\pi  &=& 1.421 \quad GeV  \nonumber \\
  \Lambda^*_\rho &=& 2.273 \quad GeV  \\
f_{N\Delta\rho}  &=& 7.4 \quad .       \nonumber
\end{eqnarray}
We have found that in this case a good fit could be obtained
already without a $\sqrt{s}$--dependence, except at the highest energy.
The results of our fits in comparison to the experimental data
\cite{Dmitriev86} at 0.970, 1.48 and 2.02 GeV are shown in Fig. 4.
Up to 2.0 GeV
we have found a destructive interference between $\pi $- and $\rho $-meson,
above 2.0 GeV there is a constructive interference.
We then conclude that now also the $N\Delta$--meson vertices are
determined; as for the case of the NN--meson vertices we assume them to be
correct also for half--off--shell--processes.

In order to check this hypothesis
we perform calculations for the pion production in nucleon-nucleon
collisions as a first test to verify our potentials.
We compare our calculations with experimental data for
five--times--differential cross
sections for $pp\rightarrow pn\pi^+$ \cite{Hudo}. Figure 5 shows
good agreement between experimental data and our calculation.

For smaller energies the fitted coupling constant are not very energy
dependent.
We thus used the coupling for 800 MeV also for smaller energies (400 MeV --
800 MeV). With these
parameters we also reproduce the total cross sections for elastic
nucleon--nucleon scattering within 20\%.

We take this as a justifaction to calculate with these parameters
also the total cross sections for the
$pp\rightarrow pn\pi^+$ process for lower energies.
Figure 6 shows that we get a good agreement with experiment
(experimental data are taken from \cite{Lan}).
Using only Feynman graphs with propagating $\Delta$ we get much too
low cross sections for small
energies; there the production over the nucleon pole is obviously
very important.

\section{DILEPTONS}

After having convinced ourselves in the last section that the effective
OBE model gives a good description of elastic scattering and
pion production amplitudes, we now use the interaction vertices determined
there for a calculation of dilepton production in elementary
nucleon--nucleon collisions. In this calculation we first neglect
any electromagnetic formfactors of the nucleons and determine
the coupling constant
$f_{N\Delta \gamma } $ from the $\gamma-$decay branch of the $\Delta$.
The vertex function for this process is:
\begin{eqnarray}
\label{NUCLEONDELTAGAMMA}
-i\frac{f_{N\Delta\gamma}}{m_\pi} \left(q^\beta\gamma_\beta g_{\mu\alpha}
 - q_\mu\gamma_\alpha\right)\gamma_5
                         \nonumber \\
  f_{N\Delta \gamma }=0.32 \quad ,
\end{eqnarray}
where $q$ is the momentum of the photon in this case. The coupling
constant has been obtained from the experimental partial width
of the $\Delta$ of 0.6\% for the decay into a real photon.

We also include the coupling to the anomalous magnetic
moment of the nucleon. A consequence of this is that now also the
neutron lines can contribute to the radiation of dileptons. Both
the $\Delta$ resonance and the radiation from neutron
lines were not included in our earlier calculations \cite{Schaefer89}
nor in those of Haglin et al. \cite{Haglin89}.
That the latter becomes more important for higher photon-energies
was already found in our calculations of real bremsstrahlung
\cite{Schaefer91}.

Figs. 7 and 8 show the invariant mass spectra for dilepton production
both for pp and pn collisions at the two bombarding energies 1.0 and 2.1
GeV. In all cases the dominant contribution arises from the resonance
amplitude where the intermediate, off-shell particle is a $\Delta$; in
fact the total yield is almost equal to the contribution of the $\Delta$
amplitude alone. The pn amplitude is for these graphs about a factor 2
-- 3 larger than for the pp reactions; this factor is mainly determined
by different isospin factors.

The graphs involving only intermediate nucleon lines are for the pn
reactions significantly larger than for pp (factors 6 and 4 at 1 and 2.1
GeV, respectively). This is due to the destructive interference between
the radiation from
both particle in the case of pp scattering. The total
yield from pn is larger than that from pp by only about a factor of 4 at
1 GeV and a factor of 2.5 at 2.1 GeV (both at $M=0.2$ GeV).

At first sight the mass-spectra of the nucleon graphs and the $\Delta$
graphs alone look very similar. However, a closer inspection shows that
that at 1 GeV both merge at the high end of the mass spectrum. In this
situation a sizeable interference of both contributions also takes
place. Note also that the increase of the bombarding energy leads to a
hardening of the spectra, but not to an increase of the value at small
$M$.

In a first comparison with experimental data we now look at the latest
results of the BEVALAC group\cite{Naudet,Wilson} for the ratio of
the pd-- to pp--dilepton yield. In Fig. 9 we show this ratio for
lab energies from 1.03 GeV up to 4.9 GeV, the experimental
data\cite{Naudet,Wilson}
and the results of our calculations (solid line).
The calculation for the deuteron was performed by summing incoherently
over the total pn and pp cross sections.
Since our model
gives the correct scattering amplitudes only up to 3.0 GeV we omit the
calculation for 4.9 GeV and show the data for completeness only.
It is seen that the calculation reproduces the general trend of the
$pd/pp$ ratio very well. In particular it is able to describe the
absolute value of the ratio as well as its mass-dependence correctly;
also the rise with $M$ towards the kinematical limit at the lower
bombarding energies is obtained, although maybe somewhat too small (at
1.26 GeV).

A closer inspection of the reasons for this behavior shows that we get
a destructive
interference between the nucleon-- and the $\Delta$--contributions for
the pp reaction which
in the high mass region is as
large as the nucleon contribution (see Fig. 7--8), the largest one there.
For the pn reaction at the kinematical limit we obtain also a destructive
interference which is as large as the
in the pp case, but for all invariant masses smaller
than the largest of the individual contributions by at least a factor 2--3 so
that it has virtually no effect in the pn case.

Figure 10 shows the results of the calculation, together with
data, for the $p + ^9Be$ reaction \cite{Naudet89}. Note that for this
comparison we have made the assumption that all dileptons are radiated
incoherently and that none of the $\Delta$'s are reabsorbed before they
can emit a dilepton pair.
This means that the cross sections from pn-- as well as pp--scattering
are summed up, weighted by the number of neutrons or protons in the
nucleus, respectively.

The efficiency of the detector is included in the integration
over the phasespace--volume by using the latest filter--routine
from the DLS collaboration\cite{Roche91}. We use the same filter
for all the various dilepton--sources.
Below 300 MeV invariant mass of the dilepton pair
the filter can decrease the yield
by factors of two or three;
in this energy range we fail to describe the data. In particular,
we cannot describe the dip in the data around 150 MeV.

While we overestimate the data at 1 GeV we obtain good agreement at 2.1
GeV, except for invariant masses
around the $\rho$--mass (see Fig. 10), where the calculated
values are somewhat too low.
Note also that the 2.1 GeV data
contain a contribution from $\eta-$Dalitz decay; we take this
contribution from ref.\cite{Wolf}.

While the resonance terms give the largest contributions for the
elementary process, their influence may be overestimated when
scaling this cross section up to the experimental case of
$p + ^9Be$, because of a possible reabsorption
of the $\Delta$ by another nucleon before emitting the virtual photon,
the Pauli--blocking of the final state and possible shadowing
effects.
A comparison of our calculations with
experiments on the elementary process $pp \rightarrow ppe^+e^-$ would be
free of such uncertainties.

\subsection{Gauge invariance}

Besides the contribution from bremsstrahlung, our calculation
also includes dilepton production from the charged internal meson lines
(Fig. 1c). For this internal radiation, from the pions, for example,  we
could use the experimentally well established
electromagnetic formfactor of the pion \cite{WeiseBuch} which then would
lead to an enhancement of the cross section just in the region
of the $\rho$--mass.
This procedure, however, is not gauge--invariant,
since current conservation requires
\begin{eqnarray}
\label{GAUGECONDITION}
   q^\mu J_\mu =  0 \quad,
\end{eqnarray}
so that
at a first sight the nucleons must have the same electromagnetic
structure as the pions. Thus we are lead to the strong VMD hypothesis,
i.e. all hadrons couple to photons by an intermediate $\rho$--meson.
The dipole shape of the spacelike electromagnetic formfactor of
the nucleon shows that this is too strong a restriction. This
section shows how to keep gauge invariance with different formfactors
for the various hadrons.

The problem of losing gauge invariance when using electromagnetic
formfactors arises because we have so far not used the complete
vertex function for the half--off--shell photon production vertices.
The full photon vertex for the nucleon is \cite{Tiemeijer90}:
\begin{eqnarray}
\label{NUCLEONPHOTONVERTEX}
  \Gamma_\mu (p^\prime ,p)& = &  \nonumber\\
     &  e \bigg( &
       \Lambda ^+(p^\prime ) \left[ F^{++}_1 \gamma _\mu +
                              F^{++}_2 \frac{i \sigma _{\mu \nu} q^\nu }
                                            {2m_N}  +
                              F^{++}_3 q_\mu
                      \right] \Lambda^ +(p)  \nonumber \\
 & + &\Lambda^+(p^\prime ) \left[ F^{+-}_1 \gamma _\mu +
                              F^{+-}_2 \frac{i \sigma _{\mu \nu} q^\nu }
                                            {2m_N}  +
                              F^{+-}_3 q_\mu
                      \right] \Lambda ^-(p)   \nonumber \\
 & +    &\Lambda^-(p^\prime ) \left[ F^{-+}_1 \gamma_ \mu +
                              F^{-+}_2 \frac{i \sigma _{\mu \nu} q^\nu }
                                            {2m_N}  +
                              F^{-+}_3 q_\mu
                      \right] \Lambda ^+(p)   \nonumber \\
&  + & \Lambda^-(p^\prime ) \left[ F^{--}_1 \gamma _\mu +
                              F^{--}_2 \frac{i \sigma _{\mu \nu} q^\nu }
                                            {2m_N}  +
                              F^{--}_3 q_\mu
                      \right] \Lambda ^-(p)  \bigg)
\end{eqnarray}
where $p$ and $p^\prime $ are the initial and final nucleon four-momenta of the
nucleon, which can be off-shell. The formfactors $F$ depend on the
three variables $p^2$,$p^{\prime 2}$ and the photon momentum squared, $q^2$.
The indices $+$ and $-$
denote the
positive and negative energy states of the nucleon, the quantities
\begin{eqnarray}
  \Lambda^\pm(p) &=& \frac {\pm p_\mu \gamma ^\mu + W} {2W} \nonumber\\
  W &=& (p^2)^{1/2}
\end{eqnarray}
are the corresponding projection operators.
By using the Ward-Takahashi identity (WTI) \cite{Itz}
\begin{eqnarray}
\label{WTI}
  q_\mu \Gamma ^\mu (p^\prime,p)&=&e e_N [S^{-1}(p^\prime )-S^{-1}(p)]
\nonumber\\
  e_N&=& (1+\tau_3)/2 ,
\end{eqnarray}
where $S(p)$ is the propagator of the particle,
we find a relation between $F_1 $ and $F_3 $:
\begin{eqnarray}
\label{F1F3RELATION}
  &\Lambda ^\pm (p^\prime )& F^{\pm \pm}_3  \Lambda ^\pm (p)= \nonumber \\
\frac {1}{q^2} &\Lambda ^\pm (p^\prime )&
  \left( -q_\mu \gamma ^\mu F^{\pm \pm}_1
        +e_N \left[S^{-1}(p^\prime )-S^{-1}(p) \right] \right)
        \Lambda ^\pm (p),
\end{eqnarray}
where $e_N$ distinguishs between protons and neutrons.
For given formfactors $F_1$ we can thus satisfy the WTI
by choosing the $F_3$ formfactors according to eq.(\ref{F1F3RELATION}).

That this will restore
gauge-invariance can be easily seen if we check eq.(\ref{GAUGECONDITION})
for pn-bremstrahlung by an exchange of an {\em uncharged}
meson. The various contributions to the four--divergence are (see Fig. 1):
\begin{eqnarray}
\label{DIAGRAM12}
   Diagram \quad 1a : & &\nonumber \\
   q_\mu J^\mu&=&\bar u(p_3) \left[ \Gamma_{NN} S(p_1+q) q_\mu \Gamma ^\mu
                             \right] u(p_1) \nonumber \\
              &\times &\bar u(p_4) \Gamma_{NN} u(p_2) D(p_4-p_2) \nonumber \\
              &=&\bar u(p_3) \left[ \Gamma_{NN} S(p_1+q) S^{-1}(p_1+q)
                             \right] u(p_1)  \nonumber \\
              &\times &      \bar u(p_4) \Gamma_{NN} u(p_2) D(p_4-p_2)
                                                       \nonumber \\
              &=&\bar u(p_3) \Gamma_{NN} u(p_1)
                 \bar u(p_4) \Gamma_{NN} u(p_2) D(p_4-p_2) \nonumber \\
   Diagram \quad 1b : \nonumber\\
   q_\mu J^\mu &=& \bar u(p_3) \left[ q_\mu \Gamma ^\mu S(p_3-q) \Gamma_{NN}
                               \right] u(p_1) \nonumber \\
                &\times &\bar u(p_4) \Gamma_{NN} u(p_2) D(p_4-p_2) \nonumber\\
                &=&\bar u(p_3) \left[ - S^{-1}(p_3-q) S(p_3-q) \Gamma_{NN}
                               \right] u(p_1)  \nonumber \\
                &\times &      \bar u(p_4) \Gamma_{NN} u(p_2) D(p_4-p_2)
                                                       \nonumber \\
              &=&- \bar u(p_3) \Gamma_{NN} u(p_1)
                   \bar u(p_4) \Gamma_{NN} u(p_2) D(p_4-p_2) \nonumber \\
                                                           \nonumber \\
   Diagram \quad 1a &+& Diagram \quad 1b = 0 \qquad   q.e.d \quad .
\end{eqnarray}
Here $D(p)$ is the meson propagator and $S(p)$ that of a nucleon;
eq. (\ref{WTI}) has been exploited in these expressions.

In the case of {\em charged} meson exchange one has to interchange $p_3$ and
$p_4$
and use in diagram 1b $D(p_1-p_4)$ instead of $D(p_4-p_2)$.
In this case the sum of both diagrams does not vanish. This is because
the meson can radiate a photon, too.
Thus we get the additional diagram 1c:
\begin{eqnarray}
\label{DIAGRAM3}
 q_\mu J^\mu&=& \bar u(p_4) \Gamma_{NN} u(p_1) \bar u(p_3) \Gamma_{NN} u(p_2)
\nonumber\\
 &\times & D(p_1-p_4) \Gamma_{(\pi)\mu} q^\mu D(p_3-p_2)\nonumber\\
 &=& \bar u(p_4) \Gamma_{NN} u(p_1) \bar u(p_3) \Gamma_{NN} u(p_2) \nonumber\\
 &\times & D(p_1-p_4) \left[ D^{-1}(p_3-p_2)-D^{-1}(p_1-p_4)\right]
D(p_3-p_2)\nonumber\\
 &=&  \bar u(p_4) \Gamma_{NN} u(p_1) \bar u(p_3) \Gamma_{NN} u(p_2) \nonumber\\
  &\times &  \left( D(p_1-p_4) - D(p_3-p_2) \right) \quad .
\end{eqnarray}
The sum of all three diagrams vanishes.

In our actual calculation the hadronic vertices contain also strong
formfactors depending on the four-momentum of the exchanged meson
(eq.\ref{STRONGFORMFACTOR}).
For uncharged mesons this means to multiply all the above diagrams simply
by the same number $F_i(p_4-p_2)$ (\ref{STRONGFORMFACTOR}), thus
keeping gauge invariance.
In the case of charged meson exchange, however, the four-momentum of
the meson
changes in diagram 1c. This change can be taken into account by an
additional factor in eq.(\ref{DIAGRAM3}) \cite{Haglin89,Mathiot}:
\begin{eqnarray}
\label{GAUGINGFORMFACTOR}
  1 + \frac{m_{meson}^2-q_1^2}{\Lambda^2-q_2^2}
    + \frac{m_{meson}^2-q_2^2}{\Lambda^2-q_1^2}
\end{eqnarray}
which can be interpreted as the radiation of photons
by the heavy cutoff-particles which carry the same quantum numbers as the
corresponding meson.

In \cite{Tiemeijer90,Gross87} the same result is obtained in a different
way.
There the strong interaction formfactors and the free meson propagator
are used together
as an effective meson propagator. Then the WTI for the meson is fulfilled
for this new propagator. This leads to electromagnetic formfactors for
the meson-gamma vertex,
which agree with eq.(\ref{GAUGINGFORMFACTOR}).
Thus from this way
of gauging the strong formfactors it is evident that there is
the possibility to use a given electromagnetic formfactor for the meson
and a different one for the nucleon and still to fulfil the WTI.

\subsection{Formfactors}

Coming back to the formfactors, it was shown in
eq.(\ref{NUCLEONPHOTONVERTEX})
that all formfactors depend on three kinematical variables. The
dependence of the on--shell formfactors on the invariant mass $M$
of the dileptons
can be obtained for the hadrons from various experiments, at least
for some specific region of $M$.

For the pion we use the electromagnetic formfactor, experimentally
well determined from
$\pi^+\pi^-$--annihilation \cite{WeiseBuch}
\begin{eqnarray}
\label{PIONFORMFACTOR}
  F_\pi (M^2) &=& \frac {m^2_V}{m^2_V - M^2} \nonumber \\
  m_V &=& 770 \quad MeV \quad .
\end{eqnarray}
This enters into the calculation of the contribution of diagram 1c.

Unfortunately, for the nucleons
information on the formfactor is only available for
$M^2 < 0$ (from electron-proton scattering) and for
$M > 2m_N$ (from $e^+e^- \leftrightarrow p\bar p$).
The region $0 \le M \le 2m_N$ is for the nucleons not accessible in
any on--shell experiment; this is unfortunate since this is just the
interesting region that includes the vector--meson masses and
is thus crucial for establishing the validity of vector meson
dominance for the nucleon.

In order to test the sensitivity of our results to the
formfactors of the nucleon we use two different models.
In the first we extrapolate the model of ref.\cite{Brown86},
which gives an excellent fit in the space-like domain,
into the timelike region.
In that model the formfactor arises from the meson cloud surrounding the
quark core as well as from the core itself.
\begin{eqnarray}
\label{NUCLEONFORMFACTORfromWEISE}
  F_C(M^2) &=& \frac{\Lambda^2}{\Lambda^2 - M^2} \nonumber\\
  F_R(M^2) &=& \frac{2 m_V^4} { (m^2_V - M^2) (2 m^2_V - M^2)} \nonumber\\
  F_1^{proton}(M^2) &=& \frac{1}{2} (F_C(M^2) + F_R(M^2)) \\
  F_2^{nucleon}(M^2) &=& F_R(M^2) \nonumber\\
  \Lambda &=& 893 \quad MeV \nonumber \\
  m_V &=& 650 \quad MeV \quad . \nonumber
\end{eqnarray}
In the actual calculation the $F_2$--formfactor is multiplied
with the proper anomalous magnetic moment for a proton or a neutron.
For the $\Delta$ we take for simplicity the same formfactor $F_2$ as
for the nucleon.

In the second model we use the same formfactor as for the
pion (\ref{PIONFORMFACTOR}) also for the nucleon and the $\Delta$.
This would correspond to strict vector meson dominance.

By adding to the masses and cutoffs a typical width of 100--150 MeV
we avoid the pole structure of these formfactors and make the latter
complex. Note, however, that this procedure does not include any
imaginary parts resulting from other possible decay channels. Our results
depend
especially in the pole region on the value we take for the width,
but if we believe in VMD this width should not be too different
from the width of the rho--meson. The formfactors for the pion
and nucleon are shown in Fig. 11. Note that the model of
ref.\cite{Brown86} exhibits a strange structure in $F_1$ and $F_2$
in the high mass region, which may simply indicate that this
model has been extrapolated beyond its limits of validity.

In order to get a feeling for the influence of the formfactors on the
spectra we now make two assumptions, namely that first these
formfactors can also be applied to the half--off--shell vertices
appearing in dilepton production without any off--shell corrections
and, second, that all the formfactors $F^{++},F^{+-}$ and
$F^{-+}$ are the same
(the $F^{--}$--formfactors do not appear in the process treated here).

The results of this calculation are shown in Fig. 12 for the
bombarding energy of 2.1 GeV. Since we are interested here only
in the effects of the electromagnetic formfactors we do not
show the $\eta-$Dalitz decay contribution (cf. Fig. 10) again.

In the region of interest, $M$ = 600 MeV or higher,
there is no contribution from the $\eta$-decay. Thus it might be possible
to extract an estimate for the hadronic
formfactors from the existing data . In Fig. 12 the dotted line corresponds to
our
calculation without
any formfactor; it underestimates the data in the mass region around
600 MeV by about a factor of 3.
Fig. 8 shows
that the contribution of the propagating $\Delta$ provides the dominant
amplitude
for this process. By using the formfactors (from ref.\cite{Brown86})
we now would overestimate the data drastically in the meson mass
region between 600 and 700 MeV (dashed line in Fig. 12).
If we use the strict VMD formfactors
we obtain the dashed curve which shows a shoulder-like structure
which is not seen in the data.

These results show that the cross section for the
production of dileptons of high
invariant mass in nucleon--nucleon collisions is very sensitive to the
electromagnetic formfactors of the nucleon and the $\Delta$.
The fact that the calculations with the formfactors included
overestimate the measured cross section in the vector meson mass region
and that the data - within their present accurracy - do not show
any indication of an enhancement in the vector meson mass region
might have several reasons. First, neglecting the half-off-shell
corrections may not be correct, second, our setting
$F^{++} = F^{-+} = F^{+-}$ may not be justified and, third,
unitarity constraints may play a role so that the
formfactors really are complex in this region.

\section{CONCLUSIONS}

In this paper have presented calculations of the cross
sections for
dilepton production in nucleon-nucleon collisions in a relativistic
scheme. As a major input for all these calculations a covariant amplitude based
on an one--boson--exchange--model was presented, which allows to describe
experimental
data for $NN \rightarrow NN$ and $NN \rightarrow N\Delta$ for lab energies
from 800 MeV up to 2.5 GeV and also works well for pion production.
We have then used the vertices so determined in a calculation of
the elementary dilepton production cross section and have discussed how
to include different electromagnetic formfactors for the hadrons while
maintaining gauge invariance whatever theoretical model for these formfactors
we insert.

In this framework we can describe the latest data on dilepton production
in pd and pp reactions, both in their absolute magnitude and in their
trends. Earlier data for $p + Be$ can also be described reasonably well,
although here discrepancies remain, mainly at the lowest energy.

We have then shown the importance of the electromagnetic formfactors of the
hadrons for the dilepton production in nucleon-nucleon and nucleon-nucleus
collisions. These reactions are a unique tool to study the
electromagnetic formfactor of the nucleons in the timelike region around
the vector-meson masses and are thus essential to check the validity of
vector-meson dominance directly, at least for the half-off-shell
formfactors.
In view of the adhoc nature of the
formfactors used here this can presently be only an explorative study which,
however, points to the need to include the half-off-shell effects of the
electromagnetic formfactors and their imaginary parts in the calculations.

Precise data on the elementary process would clearly help to put
constraints on these formfactors.
In particular
experiments are necessary where also the hadrons in the final state
are detected. Only in these experiments it might be possible to disentangle
all the various contributions coming from pre- or postemission diagrams,
from nucleon- or $\Delta$ contribution and from direct- or exchange-diagrams.

\newpage
\section*{Table}

\bigskip
\noindent

\begin{table}[h]
\begin{tabular}{c|c|c|c|c}
            & $\frac{g^2}{4\pi}$ & {\it l} &   m [GeV] &  $\Lambda$ [GeV] \\
\hline
$\pi$       & 12.562             & 0.1133     & 0.138     &  1.005 \\
$\sigma$    & 2.340              & 0.1070     & 0.550     & 1.952 \\
$\omega$    & 46.035             & 0.0985     & 0.783     & 0.984 \\
$\rho$      & 0.317              & 0.1800     & 0.770     & 1.607 \\
            & $\kappa$=6.033     &            &           &
\end{tabular}
\caption{Coupling constants used in the calculations.}
\end{table}

\newpage
\section*{Figure captions}

\bigskip
\noindent
{\bf Fig. 1} Feynman diagrams for emission of the photon after (a),
before (b) and during (c) the nucleon-nucleon interaction. The double
line denotes either an off-shell nucleon or a $\Delta$ resonance.

\bigskip
\noindent
{\bf Fig. 2} Differential cross section for proton-neutron scattering
by using only pion-exchange potential (dashed line) and by using
the contact interaction (\ref{CONTACTTERM}) in addition to the pion
(solid line).

\bigskip
\noindent
{\bf Fig. 3}
Result of our calculations for nucleon-nucleon scattering at 800 MeV,
1.7 GeV and 2.24 GeV lab
energy in comparison to experimental data\cite{Lan}. Solid line: proton-proton,
dashed line: proton-neutron. The solids dots give the pp-data while the
stars denote the pn data \cite{Lan}.

\bigskip
\noindent
{\bf Fig. 4}
Result of our calculations for the mass-differential cross section for the
reaction $pp \rightarrow n\Delta^{++}$
at 0.970, 1.5 and 2.02 GeV lab energy in comparison to experimental
data\cite{Dmitriev86}.
$\mu$ is the mass of the $\Delta$.

\bigskip
\noindent
{\bf Fig. 5}
Result of our calculations for the five-times differential cross section
for pion production in proton-proton reactions
at 800 GeV lab energy in comparison to experimental data\cite{Hudo}.
Solid line: all diagrams,
dashed line: only resonance diagrams.

\bigskip
\noindent
{\bf Fig. 6}
Result of our calculations for the cross section
of pion production in proton-proton scattering
in comparison to experimental data\cite{Lan}.

\newpage

\bigskip
\noindent
{\bf Fig. 7}
Result of our calculations for the cross section
of dilepton production in proton-nucleon scattering at 1.0 GeV .
The diamonds gives the contribution of non-resonance
diagrams; the resonance
contribution to this reaction is given by the crosses(+).
The squares and the crosses(x) gives the positive, respectively
negative interference between resonances and non--resonance amplitudes.
The coherent sum of all contributions is given by the solid line.
The individual contributions for the proton--proton reaction
are shown in a), for the proton--neutron reaction in b).

\bigskip
\noindent
{\bf Fig. 8}
Result of our calculations for the cross section
of dilepton production in proton-nucleon scattering at 2.1 GeV.
For further details see figure 7.

\bigskip
\noindent
{\bf Fig. 9}
Results of our calculations (solid lines) for the pd/pp -- ratio
of the dilepton yield for energies from 1.03 GeV to 4.9 GeV in
comparison with the experimental data from \cite{Wilson}. $M$ is the invariant
mass of the dilepton pair.

\bigskip
\noindent
{\bf Fig. 10}
Result of our calculations for the cross section of dilepton production
in p-Be scattering for 1.0 and 2.1 GeV, including an experimental
filter. The experimental data are taken from ref.\cite{Naudet89}.
For the reaction at 2.1 GeV we show our calculations with (solid line)
and without (dashed line) the eta-contributions taken from ref.\cite{Wolf}.

\bigskip
\noindent
{\bf Fig. 11}
Absolute value of the formfactors for pion (dash-dotted),
$F_1$-proton (solid) and $F_2$-proton (dashed).

\bigskip
\noindent
{\bf Fig. 12}
Effects of electromagnetic formfactors on the dilepton invariant mass spectra
in proton--Be scattering at 2.1 GeV in comparison to experimental data.
The dotted line shows the result of our full calculation without any
formfactors. For the solid line we include the electromagnetic formfactor
for the proton, pion and $\Delta$ of ref.\cite{Brown86}. The dashed curve
gives the result using one and the same formfactor
(eq. \ref{PIONFORMFACTOR}) for all
three particles. The data are from
ref.\cite{Naudet89} for the p--Be reaction.

\end{document}